\documentstyle[prl,aps]{revtex}

\begin{document}

\draft

\title{~\\~\\~\\~\\
Incompatible and contradictory retrodictions\\
in the history approach to quantum mechanics}

\author{Giulio Peruzzi and Alberto Rimini}

\address{Dipartimento di Fisica Nucleare e Teorica,
Universit\`a di Pavia, I--27100 Pavia, Italy}

%\date{}

\maketitle

\begin{abstract}
~\\
We illustrate two simple spin examples which show that in the consistent 
histories approach to quantum mechanics one can retrodict with certainty 
incompatible or contradictory propositions corresponding to non-orthogonal or, 
respectively, orthogonal projections.
\end{abstract}


\section{Introduction}

A problem in the consistent--history approach to quantum mechanics is that in 
general there are different consistent sets of histories compatible with facts 
ascertained at certain times such that, depending on the choice of the 
consistent set, one can infer (in fact retrodict with certainty) definite 
values of incompatible physical quantities at other times. This feature of the 
theory was recognized and discussed by Griffiths in his seminal 
paper~\cite{GRIFFITHS}. A simple example was given, where the considered 
system was a spin 1/2. Griffiths' answer to the problem was that incompatible 
physical quantities cannot jointly be said to have definite values at a certain 
time because, according to the principles of the consistent--history approach, 
the corresponding properties cannot all be inserted in a single consistent 
history. Some time later, d'Espagnat~\cite{DESPAGNAT} objected that such an 
attitude entails giving up one of the most important rules of logic. A type of 
answer different from that of Griffiths is the distinction proposed by 
Omn\`es~\cite{OMNES} between {\em true\/} and {\em reliable\/} propositions. The 
propositions involved in the retrodictions referred to above would not be true 
but only reliable. 

Recently, Kent~\cite{KENT} has shown that, if the Hilbert space of the system
is at least three--dimensional, one can retrodict with certainty different
properties corresponding to orthogonal projections. A particular case was 
considered previously by Aharonov and Vaidman~\cite{AV} in a conceptual 
framework different from the history approach. In a subsequent paper 
Kent~\cite{KENT2} recalls his previous result and uses it to support the 
proposal of a new criterion for selecting meaningful sets of histories, called 
{\em ordered consistency\/}, which would rule out this kind of difficulty. We 
think that points of view different from those of Griffiths and Omn\`es are 
legitimate, so that it is worthwhile exploring further the subject. 

The arguments in Refs.\ \cite{KENT} and \cite{AV} have a formal character, no 
example based on a specific physical system and specific physical quantities 
being given. It is our purpose to provide such an example, the physical system 
being a spin 1. In section 2, after briefly recalling Kent's formalism, we 
frame into it Griffiths' and our examples. In section 3 we comment on the 
results.

\section{Incompatible and contradictory retrodictions for spin systems}

In the Hilbert space $\cal H$
let $|i\rangle$ be the normalised state of a closed system at some initial
time $t=t_{0}$. Consider the set of histories
$\cal S$ specified at time $t_{1} > t_{0}$ by the properties
\begin{equation}
P^{1}_{1} = \frac{|m\rangle\langle m|} {\langle m|m\rangle}\,,~~~~~~~
P^{1}_{2} = 1 -\frac{|m\rangle\langle m|} {\langle m|m\rangle}
\end{equation}
and at time $t_{2} > t_{1}$ by the properties
\begin{equation}
P^{2}_{1} = \frac{|f\rangle\langle f|} {\langle f|f\rangle}\,,~~~~~~~
P^{2}_{2} = 1 -\frac{|f\rangle\langle f|} {\langle f|f\rangle}\,\cdot
\end{equation}
A sufficient condition in order that $\cal S$ be consistent is the (medium)
decoherence condition of Gell-Mann and Hartle~\cite{GH}, which can
be expressed as
\begin{equation}
\langle f|i\rangle\langle m|m\rangle =\langle f|m\rangle\langle m|i\rangle. 
\label{consistenza}
\end{equation}

In the example considered by Griffiths 
the system is a non--interacting spin $1/2$ in two--dimensional
$\cal H$. Let $|i\rangle = |s_{z} = + \frac{1}{2}\rangle$ and
$|f\rangle =|s_{x} = + \frac{1}{2}\rangle$. Then it is easily seen that there 
are two ways to meet the consistency condition (\ref{consistenza}) by
$|m\rangle$.

The first possibility is that the intermediate state $|m\rangle$ coincides
with the initial state of the system, i.e. $|m\rangle =|i\rangle$. Let
${\cal S}_{\rm i}$ be the consistent set of histories corresponding to this
choice. The conditional probabilities for ${\cal S}_{\rm i}$ are shown in the
following diagram:

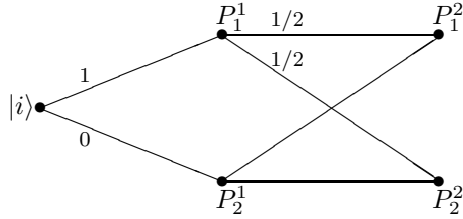
\setlength{\unitlength}{0.008in}
\begin{picture}(250,158)(-295,28)

\put(1,100){\line(5,2){119}}  
\put(1,100){\line(5,-2){119}} 
\put(120,148){\line(1,0){143}} 
\put(120,52){\line(1,0){143}}  
\put(120,148){\line(3,-2){143}}
\put(120,52){\line(3,2){143}}

\put(-20,96){$|i\rangle$}
\put(27,117){\footnotesize{$1$}}
\put(27,75){\footnotesize{$0$}}
\put(116,155){$P^{1}_{1}$}
\put(116,35){$P^{1}_{2}$}
\put(258,155){$P^{2}_{1}$}
\put(258,35){$P^{2}_{2}$}
\put(152,128){\footnotesize{$1/2$}}
\put(152,154){\footnotesize{$1/2$}}
\put(-3,96){$\bullet$}
\put(116,144){$\bullet$}
\put(258,144){$\bullet$}
\put(116,48){$\bullet$}
\put(258,48){$\bullet$}

\end{picture}

\noindent
The second possibility is that the intermediate state $|m\rangle$
coincides with the final state $|f\rangle$, i.e. $|m\rangle =|f\rangle$.
Let ${\cal S}_{\rm f}$
be the consistent set corresponding to this choice. The conditional
probabilities for ${\cal S}_{\rm f}$ are shown in the following diagram:

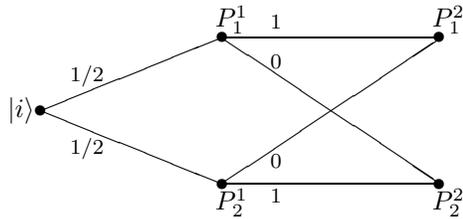
\setlength{\unitlength}{0.008in}
\begin{picture}(250,158)(-295,27)

\put(1,100){\line(5,2){119}}  
\put(1,100){\line(5,-2){119}} 
\put(120,148){\line(1,0){143}} 
\put(120,52){\line(1,0){143}}  
\put(120,148){\line(3,-2){143}}
\put(120,52){\line(3,2){143}}

\put(-20,96){$|i\rangle$}
\put(21,119){\footnotesize{$1/2$}}
\put(21,72){\footnotesize{$1/2$}}
\put(116,155){$P^{1}_{1}$}
\put(116,35){$P^{1}_{2}$}
\put(258,155){$P^{2}_{1}$}
\put(258,35){$P^{2}_{2}$}
\put(152,128){\footnotesize{$0$}}
\put(152,154){\footnotesize{$1$}}
\put(152,40){\footnotesize{$1$}}
\put(152,63){\footnotesize{$0$}}
\put(-3,96){$\bullet$}
\put(116,144){$\bullet$}
\put(258,144){$\bullet$}
\put(116,48){$\bullet$}
\put(258,48){$\bullet$}

\end{picture}

It is obvious from the diagrams that, both for ${\cal S}_{\rm i}$ and
${\cal S}_{\rm f}$, from the property
$P^{2}_{1}$ at time $t_2$ we can retrodict with certainty
the property $P^{1}_{1}$ at time $t_1$. In other words,
we can retrodict with certainty both that the value of $s_z$ was
$+ \frac{1}{2}$ and that the value of $s_x$ was $+ \frac{1}{2}$.
We say that we are faced with {\em incompatible retrodictions}.\footnote{Kent
uses in this case the term ``complementary''.
We feel that the common meaning of the term ``incompatible'' is more
adherent to describe the situation.}

In our example the system is a non--interacting spin~$1$ in
three--dimensional $\cal H$. Let the initial state of the system be
\begin{equation}
|i \rangle = |s_{z} = 0 \rangle = \left( \begin{array}{c}
0 \\ \noalign{\vskip 2pt}
1 \\ \noalign{\vskip 2pt}
0
\end{array} \right) \label{iniziale}
\end{equation}
and the (non--normalised) state vectors $| f\rangle$ and $|m \rangle$ be
specified by
\begin{equation}
|f \rangle = |s_{\bf n} = 0 \rangle = \left( \begin{array}{c}
{\displaystyle -\frac{1}{\sqrt{2}} \frac{n_{x} -i n_{y}}{n_{z}}}\\
\noalign{\vskip 2pt}
1\\ \noalign{\vskip 2pt}
{\displaystyle \frac{1}{\sqrt{2}} \frac{n_{x} + i n_{y}}{n_{z}}}
\end{array} \right), \label{finale}
\end{equation}

\bigskip
\begin{equation}
|m_{\pm} \rangle = |s_{\bf m} = \pm 1 \rangle =  \left( \begin{array}{c}
{\displaystyle \pm \frac{1}{\sqrt{2}} \frac{m_{x} -i m_{y}}{1 \mp m_{z}}}\\
\noalign{\vskip 2pt}
1\\ \noalign{\vskip 2pt}
{\displaystyle \pm \frac{1}{\sqrt{2}} \frac{m_{x} + i m_{y}}{1 \pm m_{z}}}
\end{array} \right), \label{intermedio}
\end{equation}
in terms of two unit vectors ${\bf n}$ and ${\bf m}$ in the physical space.
The components of ${\bf n}$ and ${\bf m}$ must satisfy the conditions
\begin{equation}
n_{x}^{2}+n_{y}^{2}+n_{z}^{2}=1,\;\;  \;\;\; \;\;
m_{x}^{2}+m_{y}^{2}+m_{z}^{2}=1. \label{componentiversori}
\end{equation}
On the other hand, the consistency condition (\ref{consistenza}) reduces to
\begin{equation}
\langle m_\pm |m_\pm \rangle =\langle f|m_\pm \rangle. \label{consistenzatreD}
\end{equation}
Let us take
\begin{equation}
m_x = a n_x \, , \;\; m_y = a n_y \, ,\;\;  m_z = - b n_z \, ,
\label{assunzioni}
\end{equation}
where $a$ and $b$ are positive real numbers.
Then the normalisation condition (\ref{componentiversori}) for ${\bf m}$
gives
\begin{equation}
a ^2 = \frac{1 - b^2 n_z ^2}{1 - n_z ^2}\,\cdot \label{mversore}
\end{equation}
Using assumption (\ref{assunzioni}) in expression (\ref{intermedio}) one
finds
\begin{equation}
\langle m_+ |m_+ \rangle = \langle m_- |m_- \rangle =
\frac{a^2 \big(1 - n_z ^2\big) \big(1+ b^2 n_z ^2\big)}{\big(1 - b^2 n_z ^2
\big)^2} +1 \label{mm}
\end{equation}
and
\begin{equation}
\langle f |m_+ \rangle = \langle f |m_- \rangle =
\frac {a b \big(1 - n_z ^2\big)}{\big(1 - b^2 n_z ^2\big)} +1\,. \label{fm}
\end{equation}
Thus the consistency condition (\ref{consistenzatreD}) assumes, both for
the upper and the lower sign, the form
\begin{equation}
a \big(1 + b^2 n_z ^2\big) = b \big(1 - b^2 n_z ^2\big),  \label{primaeq}
\end{equation}
i.e., taking into account Eq.\ (\ref{mversore}),
\begin{equation}
n_z ^2 = \frac{b ^2 - 1}{b ^2 ( b ^2 + 3)}\,\cdot \label{eqfinale}
\end{equation}
It is easily seen that for $b^2 \geq 1$ the value of  $n_z ^2$ is in
the interval $[0,1/9]$ and is therefore acceptable (there are two values
of $b^2$ for each value of $n_z ^2 < 1/9$).

From the above discussion it follows that, for each acceptable pair of unit
vectors ${\bf n}$, ${\bf m}$, one can consider two consistent sets of
histories ${\cal S}_+$ and ${\cal S}_-$ corresponding to $|m \rangle =
|m_+ \rangle$ and $|m \rangle = |m_- \rangle$, respectively. Chosen ${\bf n}$ 
and ${\bf m}$, the conditional probabilities are the same for ${\cal S}_+$ and
${\cal S}_-$ and are shown in the following diagram:

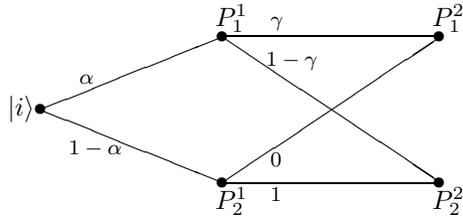
\setlength{\unitlength}{0.008in}
\begin{picture}(250,158)(-295,27)

\put(1,100){\line(5,2){119}}  
\put(1,100){\line(5,-2){119}} 
\put(120,148){\line(1,0){143}} 
\put(120,52){\line(1,0){143}}  
\put(120,148){\line(3,-2){143}}
\put(120,52){\line(3,2){143}}

\put(-20,96){$|i\rangle$}
\put(27,117){\footnotesize{$\alpha$}}
\put(20,71){\footnotesize{$1- \alpha$}}
\put(116,155){$P^{1}_{1}$}
\put(116,35){$P^{1}_{2}$}
\put(258,155){$P^{2}_{1}$}
\put(258,35){$P^{2}_{2}$}
\put(149,129){\footnotesize{$1- \gamma$}}
\put(152,155){\footnotesize{$\gamma$}}
\put(152,40){\footnotesize{$1$}}
\put(152,63){\footnotesize{$0$}}
\put(-3,96){$\bullet$}
\put(116,144){$\bullet$}
\put(258,144){$\bullet$}
\put(116,48){$\bullet$}
\put(258,48){$\bullet$}
\end{picture}

\noindent
where $\gamma = \beta / \alpha$ and
\begin{equation}
\alpha = \frac{\langle i | m \rangle \langle m | i \rangle}{\langle m | m 
\rangle} = \frac{1}{\langle m | m \rangle} =
\frac{1- b^2 n_z ^2}{2}\, \hbox{\raise 1.5pt \hbox{,}}
\end{equation}
\begin{equation}
\beta = \frac{\langle i | f \rangle \langle f | i \rangle}{\langle f | f 
\rangle} = \frac{1}{\langle f | f \rangle} = n_z ^2 \,.
\end{equation}

Let us exclude the choices $n_z=0$, $m_z=0$ (corresponding to
the solution $b=1$, $n_z=0$ of eq. (\ref{eqfinale})) and $n_z=0$, $m_z=-1$
(corresponding to $b \rightarrow \infty$, $bn_z \rightarrow 1$), for which
the probability of the property $P_1 ^2$ at time $t_2$ is zero. Then it
is apparent from the diagram that from the property $P_1 ^2$ at time $t_2$
one can retrodict with certainty both that the value of $s_{\bf m}$ was $+1$
and that it was $-1$ at time $t_1$. We say that we are faced
with {\em contradictory retrodictions}.\footnote{Kent uses
in this case the term ``contrary''. We think that the term ``contradictory''
more explicitly describes the situation. What Kent designates as
``contradictory'' could be called, in our terminology, ``exhaustively
contradictory''.} 

\section{Conclusions}

The spin--1 example given in section 2 shows that the situation formally
discussed by Kent can take place in the case of realistic observations
of real experiments. Moreover the fact that the contradictory propositions
correspond to opposite values of the same quantity $s_{\bf m}$ makes the
example particularly striking.

Various types of proposal to answer the problems related to the existence of 
incompatible and contradictory retrodictions in the history approach have 
already been discussed by Griffiths~\cite{GRIFFITHS,GRIFFITHS2},
d'Espagnat~\cite{DESPAGNAT}, Omn\`es~\cite{OMNES}, and Kent~\cite{KENT,KENT2}.
We do not enter into the details of this debate. We only point out that in
Kent's second paper, the occurrence of the difficulties is related to the fact
that consistent sets like ${\cal S}_+$ and ${\cal S}_-$ do not satisfy a
criterion, called ordered consistency, which would ensure that, in the
framework of the history approach, implication by subspace inclusion is valid.
Our realistic example gives answers to some questions raised by Kent. First, 
as already noted, it proves that Kent's situation can take place in the case 
of real experiments. Second, the two non--ordered consistent sets of histories
${\cal S}_+$ and ${\cal S}_-$ of section 2 can be considered as
coarse-grainings of two (!){\hskip -2pt} usual quasiclassical domains, the
quasiclassical variables being the same in the two cases. The decoherence of
the spin variables is due to the fact that their values are strictly correlated
to suitable ranges of such quasiclassical variables. It follows that one can
answer negatively to the question ``whether quasiclassical domains should
generally be expected to be ordered consistent sets''. Therefore one can
conclude that the constraint of quasiclassicality on consistent sets of
histories does not include that of ordered consistency, so that the latter is
actually an effective additional criterion.

\acknowledgments

This work is supported in part by the Istituto Nazionale di Fisica Nucleare.

\end{document}